\documentclass[pre,twocolumn,showpacs]{revtex4}

\usepackage{graphicx}%
\usepackage{dcolumn}
\usepackage{amsmath}

\makeatletter
\def\btt#1{\texttt{\@backslashchar#1}}%
\DeclareRobustCommand\bblash{\btt{\@backslashchar}}%
\makeatother

\begin{document}


\title{$q$-exponential, Weibull, and $q$-Weibull distributions: an empirical analysis}
\author{S. Picoli Jr., R. S. Mendes, and L. C. Malacarne}
\email{lcmala@dfi.uem.br}
\affiliation{Departamento de F\'\i sica, Universidade Estadual de Maring\'a, \\
Avenida Colombo 5790, 87020-900 Maring\'a-Paran\'a, Brazil}

\date{\today}

\begin{abstract}
In a comparative study, the $q$-exponential and Weibull
distributions are employed to investigate frequency distributions
of basketball baskets, cyclone victims, brand-name drugs by retail
sales, and highway length. In order to analyze the intermediate
cases,   a distribution, the $q$-Weibull one, which interpolates
the $q$-exponential and Weibull ones, is introduced. It is
verified that the basketball baskets distribution is well
described by a $q$-exponential, whereas the cyclone victims and
brand-name drugs by retail sales ones are better adjusted by a
Weibull distribution. On the other hand, for  highway length the
$q$-exponential and Weibull distributions do not give satisfactory
adjustment, being necessary to employ the $q$-Weibull
distribution. Furthermore, the introduction of this interpolating
distribution gives an illumination from the point of view of the
stretched exponential against inverse power law ($q$-exponential
with $q > 1$) controversy.
\end{abstract}

 \maketitle

\section{Introduction}
\label{intro}

Frequency distributions have played an important part in many
investigations in very different contexts. For example,
distribution of income\cite{pareto},  frequency of words in a long
text\cite{zipf}, forest fires\cite{5}, scientific
citations\cite{redner}, world-wide web surfing\cite{8},
ecology\cite{8a}, solar flares\cite{8b}, population distribution
in large cities\cite{pop}, economic index\cite{9}, epidemics in
isolated populations\cite{Rhodes}, among others. In these cases
power law distributions, $p_p(x)=p_0 \; x^{-\alpha}$ with $\alpha
>1$, have been identified.
In general, this distribution can not be used for arbitrary $x$
values, since the number of events described by a power law is not
finite when all positive $x$ values are considered. Thus, it is
natural to investigate deviations from a power law, which can be
identified as a curvature in log-log graphics.

One of the simplest generalizations of a power law which leads
$\int_0^{\infty} p(x) dx$ to a finite value is the Zipf-Mandelbrot
law, $p_{zm}(x)=p_0/(1+s x)^{\alpha}$ $(s>0)$, which was employed
by Mandelbrot to investigate frequency of words in long
text\cite{mandelbrot}. More recently, this distribution was
generalized in order to incorporate negative $\alpha$ values, {\it
i.e.},
\begin{equation}
\label{eq1} p_q(x)=\frac{p_0}{x_0} ~ \exp_q
\left(-\frac{x}{x_0}\right),
\end{equation}
where $\exp_q (-x)\equiv [1-(1-q)x]^{1/(1-q)}$ if $1-(1-q)x\geq0$
and $\exp_q (-x)\equiv 0$ if $1-(1-q)x<0$. Thus, for $q>1$ we have
a long-tailed function since $p_q(x)\propto x^{-1/(q-1)}$ for
$x\gg x_0/(q-1)$; for $q<1$ we obtain a short-tailed function
since $p_q(x)=0$ for $x\geq x_0/(1-q)$; and for $q=1$ we have an
exponential since $\lim_{q \rightarrow 1} \exp_q(-x) = \exp(-x)$.
Note also that a convenient identification of the parameters in
Eq. (\ref{eq1}) immediately leads  to the Zipf-Mandelbrot law when
$q>1$. This distribution for $q>1$ has been employed, for
instance, in connection with scientific citations\cite{tsallis3},
goal distribution\cite{malacarne}, and population distribution in
a country\cite{malacarne2}.

Eq. (\ref{eq1}) can be viewed as a generalization of the
exponential function since it basically replaces the exponential
one in the canonical ensemble of the Tsallis
statistics{\cite{tsallis,curado}}. More precisely, $p_q(x)$ can be
obtained from a maximum entropic principle. These facts motivate
us to refer to $p_q(x)$ as $q$-exponential distribution. We remark
that the Tsallis statistics has been employed in connection with
several anomalous systems  (for a recent review see Ref.
\cite{review}).

The Weibull distribution
\begin{equation}
\label{eq2} p_w(x)=p_0 ~ \frac{ r ~ x^{r-1}}{x_0^r}~
\exp\left[-\left(\frac{x}{x_0}\right)^r\right]
\end{equation}
has been employed to accomplish a systematic curvature in a
log-log graphics. The parameter $r$ is taken positive and when
$r=1$ Eq. (\ref{eq2}) reduces to an exponential. The distribution
(\ref{eq2}) was firstly employed  in a statistical theory of the
strength of material\cite{weibull}, and has been used in  many
contexts such as dielectric failure\cite{weibull1}, cardiac
contraction\cite{weibull2}, aspects of road-accident
death\cite{12b}, wind speed\cite{weibull3}, profile of foliage
area\cite{weibull4}, porcelain strength\cite{weibull5}, among
others. In particular, this distribution has been related with
multiplicative process by Laherrere and Sornette{\cite{sornette}},
who refers to $p_w(x)$ as stretched exponential.

It must be stressed that the parameters $x_0$ and $r$ control the
tail behavior of $p_w(x)$ as well as $x_0$ and $q$ dictate the
$p_q(x)$ shape. Furthermore, if we restrict our analysis to a
finite range for $x$ it is very common that $p_w(x)$ and $p_q(x)$
indistinctly adjust  a restrict set of data very well. Thus, it is
natural to investigate whether distribution, $p_w(x)$ or $p_q(x)$,
gives a better adjustment for a large set of data. This work is
addressed to investigate this question by focusing the frequency
distributions of basketball baskets in a  championship, tropical
cyclone victims, brand-name drugs by retail sales, and highway
length. In particular, for the highway length distribution neither
the $q$-exponential nor the Weibull ones lead to a satisfactory
adjustment. In order to accommodate this case and the others in a
unified framework, we introduce a distribution which smoothly
interpolates the $q$-exponential and the Weibull ones. Moreover,
this approach gives an illumination from the point of view of the
stretched exponential against inverse power law ($q$-exponential
with $q > 1$) controversy.

\section{$q$-Weibull distribution}
\label{sec:1} We refer to this interpolating function as
$q$-Weibull distribution, and it is given by
\begin{equation}
\label{eq3}  p_{qw}(x)=p_0 ~ \frac{ r ~ x^{r-1}}{x_0^r}~
\exp_q\left[-\left(\frac{x}{x_0}\right)^r\right] .
\end{equation}
Note that for $r=1$ and $q\neq 1$ we recover the $q$-exponential.
Analogously, in the limit $q \rightarrow 1$ with $r \neq 1$ or
$r=1$, we obtain  the Weibull or the exponential distributions,
respectively. Furthermore, when $q>1$ and $x\gg x_0/(q-1)^{1/r}$,
we verify that $p_{qw}(x) \propto x^{-u}$ with
$u=r[(2-q)/(q-1)]+1$. In contrast, when $q<1$ and $x \geq
x_0/(1-q)^{1/r}$, we have $p_{qw}(x)=0$.

In order to obtain a sufficiently smooth curve, it is a common
practice to employ the cumulative distribution
\begin{equation}
\label{eq4} R(x)=\int_x^{\infty} p(y) dy
\end{equation}
to investigate the frequency distribution  for a set of data. In
our case, we obtain
\begin{equation}
\label{eqa7} R_{qw}(x)= p^{\prime}_0 ~
\exp_{q^{\prime}}\left[-\left(\frac{x}{x_0^{\prime}}\right)^r\right],
\end{equation}
with $q^{\prime}=1/(2-q)$, $x_0^{\prime}=x_0/(2-q)^{1/r}$, and
$p^{\prime}_0 = p_0/(2-q)$. It must be emphasized that $R_{qw}(x)$
exists only if $q<2$. Moreover, from our construction, the
cumulative distributions, $R_{q}(x)$  or $R_{w}(x)$, follow from
Eq. (\ref{eqa7}) by taking the appropriate limits ($r\rightarrow
1$ or $q\rightarrow 1$). Therefore, the cumulative distribution of
a power law is a power again, of a $q$-exponential is another
$q$-exponential, of a Weibull distribution is a stretched
exponential, and of a $q$-Weibull is a $q$-stretched.

\begin{figure}
\centering
\includegraphics*[width=9cm, height=6cm,trim=2cm 0.5cm 1cm 1cm]{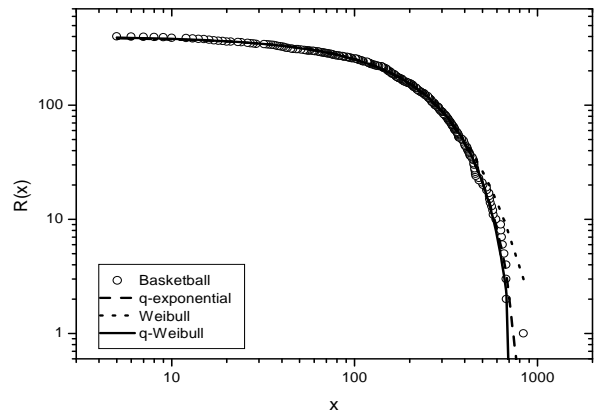}
\caption{Basketball basket cumulative distribution for two-point
shots in 1999 NBA championship. The parameters employed in $R(x)$
[Eq. (\ref{eqa7})] are obtained from those presented in Tab.
(\ref{tab1}). } \label{fig1}
\end{figure}

Before  starting our empirical investigation of data by using
these distributions, we would like  to remark  that the functions
$R_q(x)$, $R_w(x)$, and $R_{qw}(x)$ can be  related with diffusion
equations. In fact, $R_q(x)$ satisfies the generalized decay
equation $dR_q(x)/dx = - a [R_q(x)]^q$ with
$a=(1-q^{\prime}){p_0^{\prime}}^{1-q^{\prime}}/x_0^{\prime}$;
$R_w(x)$ is intimately related with the point source solution of
the anomalous diffusion proposed by O'Shaughnessy and
Procaccia{\cite{procaccia}; and $R_{qw}(x)$ occurs in the solution
of the nonlinear diffusion equation proposed in Ref.
\cite{malacarne3}. Moreover, $R_{qw}(x)$ obeys the generalized
decay equation $dR_{qw}(x)/dx = - b~x^{r-1} [R_{qw}(x)]^{q´}$ with
$b=r(1-q^{\prime}){p_0^{\prime}}^{1-q^{\prime}}/{x_0^{\prime}}^r$.

\begin{figure}
\centering
 \includegraphics*[width=9cm, height=6cm,trim=2cm 0.5cm 1cm 1cm]{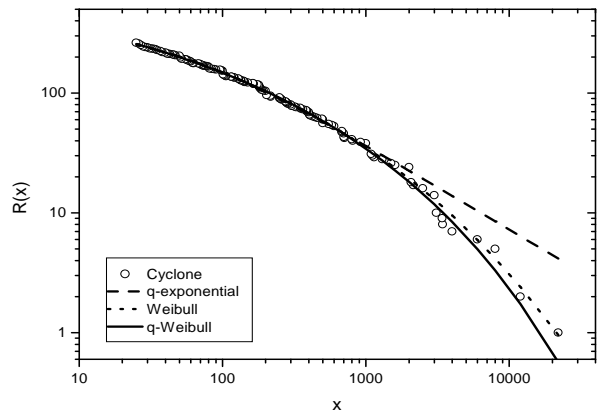}
\caption{Fatal victims cumulative distribution for tropical
Atlantic cyclones. The parameters employed in $R(x)$ [Eq.
(\ref{eqa7})] are obtained from  those presented in Tab.
(\ref{tab1}). } \label{fig2}
\end{figure}
\section{Empirical Analysis}

In this section, we are going to apply the $q$-exponential,
Weibull, and $q$-Weibull distributions to analyze the data about
basketball baskets in a championship, tropical cyclone victims,
brand-name drugs by retail sales, and highway length. To obtain
the best set of values for the parameters, we use the least square
minimum method. Complementary, as an additional tool in the
analysis of how good the adjustment is, we use the residual error
analysis. In this kind of analysis, a good adjustment is related
to a gaussian residual frequency distribution and a random
distribution of errors.


\begin{figure}[t]
\centering
\includegraphics*[width=9cm, height=6cm,trim=2cm 0.5cm 1cm 1cm]{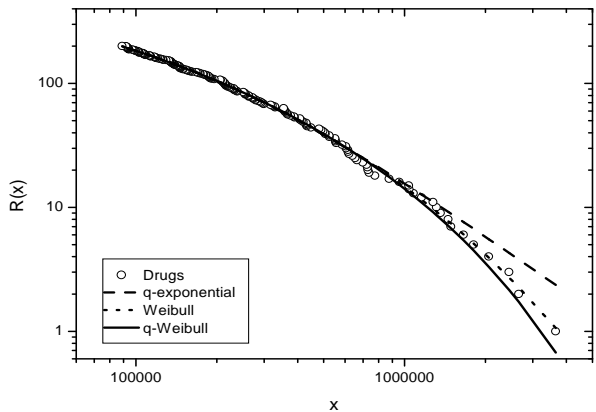}
\caption{Cumulative distribution of brand-name drugs by retail
 sales. The parameters employed  in $R(x)$ [Eq. (\ref{eqa7})] are
 obtained from  those presented in Tab. (\ref{tab1}).}\label{fig3}
\end{figure}

Naturally, since the $q$-Weibull distribution the
$q$-expo\-nential or the Weibull ones  has as limit case; in
general, it gives a better adjustment than the two others.
However, there are cases where the adjustment with the $q$-Weibull
distribution does not give a significative improvement when
compared with that obtained from one of the two other
distributions. In such cases, either the parameter $r$  is close
to one leading  to a $q$-exponential distribution or the parameter
$q$  is close to one leading to a Weibull distribution. As we are
going to verify below, there are cases where such limits are not a
good approximation, therefore indicating that the $q$-Weibull
distribution  gives a significative improvement in the adjustment.
Now we describe the systems to be analyzed.

\begin{figure}
\centering
\includegraphics*[width=9cm, height=6cm,trim=2cm 0.5cm 1cm 1cm]{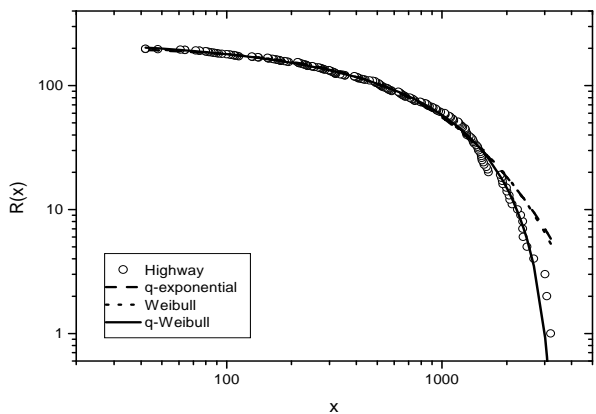}
\caption{Length highway cumulative distributions in USA. The
parameters employed in $R(x)$  [Eq. (\ref{eqa7})] are obtained
from those presented in Tab. (\ref{tab1}). } \label{fig4}
\end{figure}

It is natural to investigate aspects related to sports since it
has attracted interest of a large number of people in our society.
In fact, many studies concerning  sports have been developed. For
instance, those related to frequency distributions in
soccer\cite{malacarne}, baseball\cite{rosner,mosteller,gill},
golf\cite{mosteller}, ho\-ckey\cite{mosteller,gill},
football\cite{mosteller,gill}, and
basketball\cite{mosteller,vollmer}. Here, we concentrate our
attention on basketball. More precisely, we consider the
distribution of two-point basket among approximately  four hundred
players of 1999 NBA championship\cite{basquetedados}.

In our analysis about basketball baskets, we verified that the
$q$-exponential function gives a better adjustment than the
Weibull distribution since the parameter $r$, obtained from the
$q$-Weibull one,  is close to one $(r=0.91)$. The basketball
basket cumulative distribution, as well as the corresponding fits,
is presented in Fig. (\ref{fig1}). Tab. (\ref{tab1}) contains the
parameters employed in the fits presented in Fig. (\ref{fig1}). We
remark that in the previous examples about $q$-exponential
distribution (Refs. \cite{tsallis3,malacarne,malacarne2}) the
parameter $q$ is bigger than one, in contrast with the present
example where $q=0.68$. As far as we are concerned  this is the
first system whose adjustment leads to a $q$-exponential
distribution with $q<1$. Thus, our adjustment suggests the
existence of a cutoff for the maximum number of baskets $x_c$ for
1999 NBA championship. This $x_c= x_0^{\prime}/(1-q^{\prime})=9.9
\times 10^2$  is compatible with the true maximum number of
baskets, 839. In contrast to this case, now we focus on systems,
which are well described by a Weibull distribution.


\begin{table}
\caption{Fit parameters  to $q$-exponential, Weibull, and
$q$-Weibull distributions for the analyzed systems.}\label{tab1}
\begin{tabular}{|c|c|c|c|c|}
  \hline System & & \,$q$-exponential\, & Weibull & \, $q$-Weibull\,\\
  \hline
   Basketball & \, q \,& 0.68 & - & 0.51\\
   & $r$ & - & 1.15 & 0.91\\
   & $x_0$& 313 & 212 & 387\\
   & $p_0$& 520 & 387 & 601\\
\hline
   Cyclone & q& 1.58 & - & 0.95\\
   & $r$ & - & 0.23 & 0.22\\
   & $x_0$& 24.9 & 4.41 & 6.39\\
   & $p_0$& 145 & $1.14\times 10^3$ & $1.18\times 10^3$\\
\hline
   Drug & q& 1.39 & - & 0.88\\
   & $r$ & - & 0.35 & 0.25\\
   & $x_0$&  & $1.30\times 10^4$ & $7.03\times 10^3$\\
   & $p_0$& 266 & $1.41\times 10^3$ & $2.49\times 10^3$\\
\hline
   Highway & q& 1.12 & - & 0.077\\
   & $r$ & - & 0.87 & 0.57\\
   & $x_0$& 608 & 712 & $2.96\times 10^3$\\
   & $p_0$& 181 & 215 & 465
\\ \hline
\end{tabular}
\end{table}


Sudden inundations, such as those due to cyclones, are known as
the most significative natural catastrophe\cite{catastrofe},
leading along the years to massive death and loss of property.
Studies related with such phenomena enclose the spatial and
temporal distribution of cyclone for a given region\cite{regiao}.
Furthermore, attempts to identify the climatologic process that
leads to cyclone formation have employed meteorologic
satellites\cite{satelite}. However, since such technology is only
recently disposable, it is difficult to investigate cyclones that
have occurred for a long time. Thus, in order to analyze at least
in part the effects of cyclones, including those occurred  many
years ago, we consider the Atlantic tropical fatal victims number
from 1492 to 1999\cite{1492}. In this period, two hundred and
sixty-three Atlantic tropical cyclones with deaths  were
catalogued.

In the analysis of cumulative distribution of deaths as a
consequence of tropical cyclones, the adjustment with a
$q$-Weibull distribution leads to $q$ close to one ($q=0.95$).
This indicates that Weibull distribution can be in a good
agreement with the data. In Fig. (\ref{fig2}), we plot the
cumulative distribution of cyclone fatal victims versus the number
of deaths related to it by using the three distributions employed
in this work. In Tab. (\ref{tab1}), we show the correspondent
parameters.

An important part of the expense in our society is invested in
health. In this context, the use of pharmaceutic services plays a
special role. This decisively contributes to an enormous
competition among pharmaceutic industries, motivating the
appearance of new drugs every year. For these reasons, studies
have been performed, for instance, in the relationship between
pharmaceutical expenditures and income\cite{huttin}. In this work,
we investigate another aspect, the distribution of brand-name
drugs.  Here we are considering the top two hundred brand-name
drugs by retail sales in 1999 in USA\cite{dt}. As the above
system,  the adjustment with $q$-Weibull gives $q$ close to one
($q=0.88$), so the Weibull distribution leads to a better
adjustment than the $q$-exponential one. This fact can be
visualized in Fig. (\ref{fig3}) and the optimal parameters are
presented in Tab. (\ref{tab1}).

In the above systems, the $q$-Weibull distribution does not give a
significative improvement since the systems can be well described
by $q$-exponential or Weibull ones. But  this may not occur in
general. This is just the case of the following example.

In general, roads strategically connect different parts of a
country. In particular,  understanding  the myriad of aspects
related with road transport is of great value since the economy of
a region is strongly connected with the transport infrastructure.
Possible aspects related to road transport are the   study of
mortality and the morbidity due to road
accidents\cite{12b,thomas1} and the influence of the length of the
road segment in the statistical description of accident counts and
density\cite{thomas2}. Here, we focus attention on the
distribution of the highway length, by analyzing one hundred and
ninety-eight USA highways\cite{roads}. In Fig. (\ref{fig4}), we
can see that neither $q$-exponential nor Weibull distributions are
able to give a satisfactory adjustment. In this system, the
interpolating $q$-Weibull distribution is necessary to have a good
fit. In fact, the parameters $r$ and $q$ are not close to one
($r=0,57$ and $q^{\prime}=0.077$).

In every system, we had applied  the mean square minimum method to
obtain the optimal parameters. We also used the residual errors
analysis to confirm the goodness of the adjustments. To exemplify
such procedure, we present the residual analysis to highway length
distribution in Fig. (\ref{fig5}). The graphic of frequency
distribution of residues and the graphic of predicted versus
residual values to $q$-Weibull distribution [Fig. (\ref{fig5}-c)]
are better  than the corresponding ones referent to
$q$-exponential [Fig. (\ref{fig5}-a)] and Weibull distributions
[Fig. (\ref{fig5}-b)]. In fact, the histogram for the residual
errors  seems closer to a gaussian for $q$-Weibull distribution
than those ones to $q$-exponential  and Weibull distributions. The
analogous improvement occurs in the randomness observed in the
graphics of predicted versus residual values.

An alternative distribution that generalizes the $q$-exponential
one is the distribution that arrives from the differential
equation\cite{Proteinas}
\begin{equation}\label{proteina}
    \frac{d p}{dx}= - \mu p^r - (\lambda -\mu) p^q ,
\end{equation}
where $q$, $r$, $\mu$, $\lambda$ and the initial condition $p_0$
are real parameters to be adjusted.   The $q$-exponential
distribution is recovered if $\mu=0$ or if $r=q$. However, $p(x)$
obtained from Eq. (\ref{proteina}) does not contain  the Weibull
distribution as particular case,  desirable feature since, as
emphasized in this work, many systems are well adjusted by the
Weibull distribution. The fact that Eq. (\ref{proteina}) does not
contain the Weibull one as a particular case is clear since it
satisfies the differential equation
\begin{equation}\label{difWeibull}
 \frac{d p}{dx} = \left[ \frac{(c-1)}{x}  - \frac{c x^{c-1}}{x_0^c}\right]p.
\end{equation}
\begin{widetext}
In addition, we mention some  practical difficulties in the full
application of a distribution dictated by Eq. (\ref{proteina}).
First, Eq. (\ref{proteina}) does not have analytical solutions for
all $r$ and $q$, so we need to employ a numerical approach.
Second, it contains five parameters to be adjusted (one more than
$q$-Weibull). Due to these facts, the analysis and adjustment
become cumbersome. On the other hand, when the distribution
exhibits two slopes, the solution of Eq. (\ref{proteina}) seems
appropriate since neither q-exponential nor q-Weibull is able to
reproduce this behavior. This characteristic has been  explored in
the study of folded proteins\cite{Proteinas}. Another possibility
of application of Eq. (\ref{proteina}) occurs when the
distribution presents a well definite slope in the intermediate
region. This situation can be verified in linguistics, when
investigating deviation from the Zipf-Mandelbrot
law\cite{Montemurro}. Beside these difficulties (five parameters
and numerical solution of Eq. (\ref{proteina})), we have
investigated the adjustment of our data by using $p(x)$ from Eq.
(\ref{proteina}). We mention first that standard programs of
adjustment can not be used since we do not have an analytical form
to the distribution from Eq. (\ref{proteina}).  Furthermore, due
to the large number of parameters,  a method that not only does
not strongly depend on initial set of parameters but also presents
fast convergence is necessary. In this direction, we applied the
simulated annealing method, as employed in Ref. \cite{Penna}, to
obtain a good set of parameters. However, even this method was not
good enough to lead to a satisfactory result.
\begin{figure}[h]
\centering
\centering \DeclareGraphicsRule{ps}{eps}{*}{}
\includegraphics*[width=14cm, height=5cm,trim=0cm 0cm 1cm 1cm]{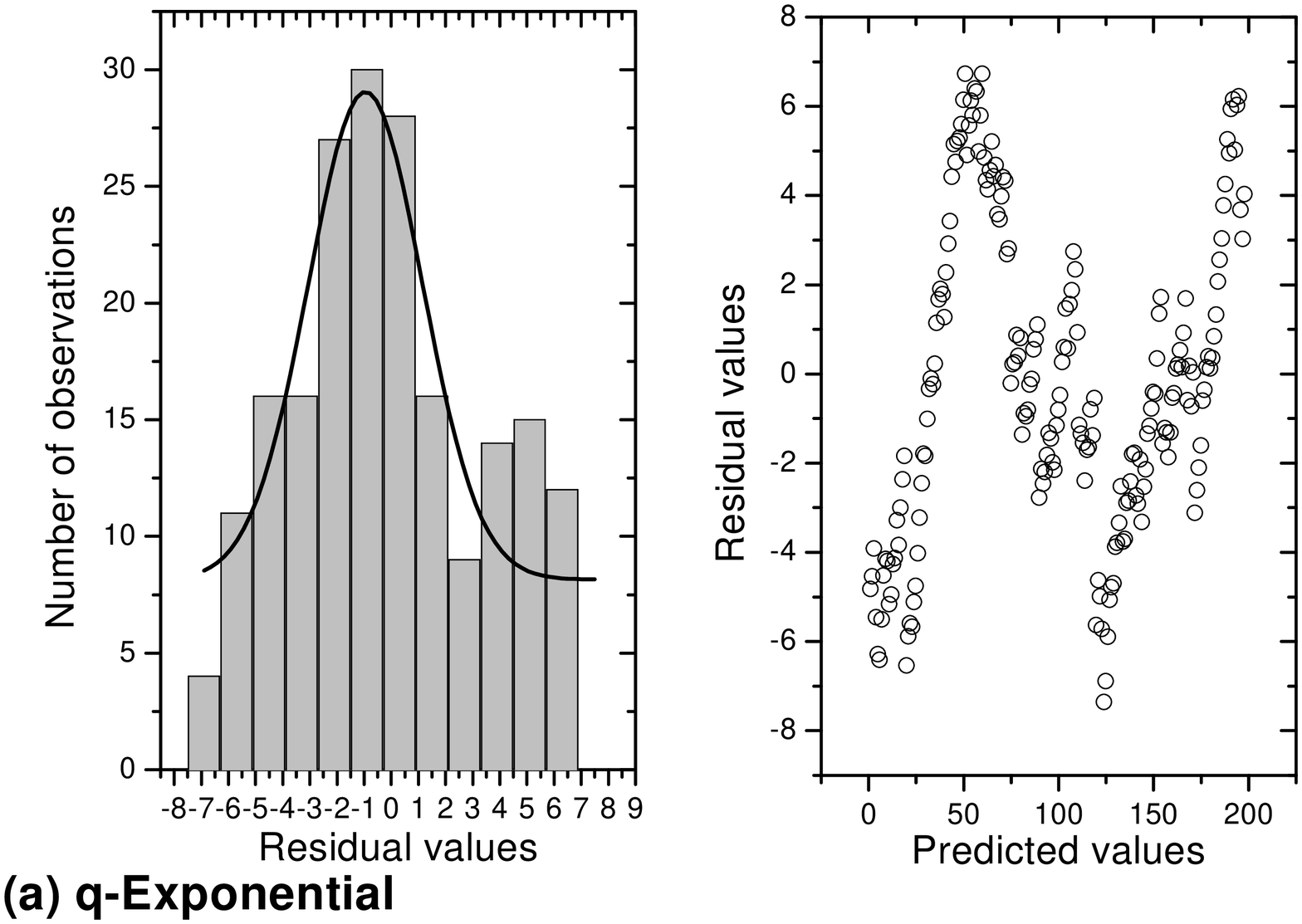}
\includegraphics*[width=14cm, height=5cm,trim=0cm 0cm 1cm 1cm]{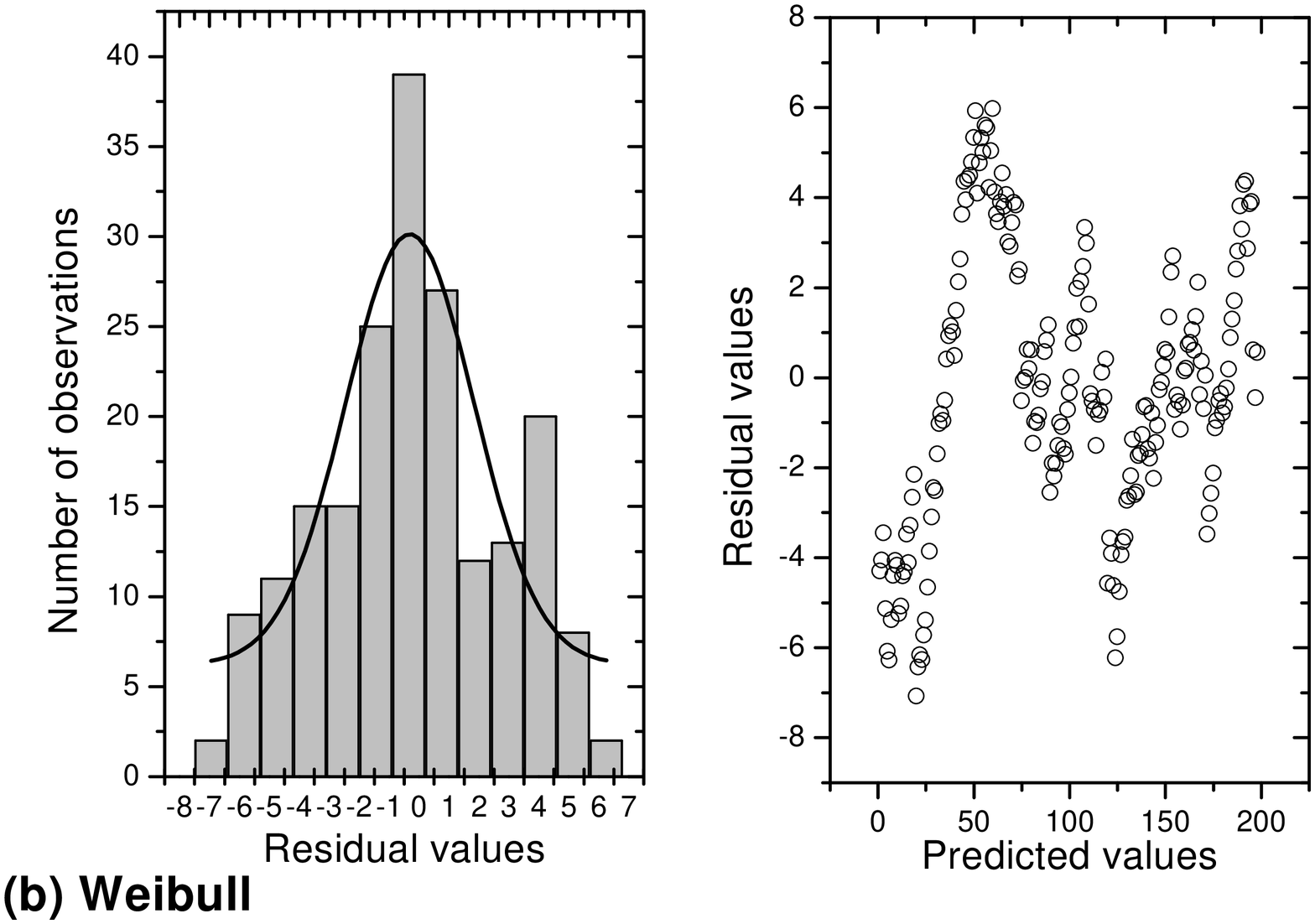}
\includegraphics*[width=14cm, height=5cm,trim=0cm 0cm 1cm 1cm]{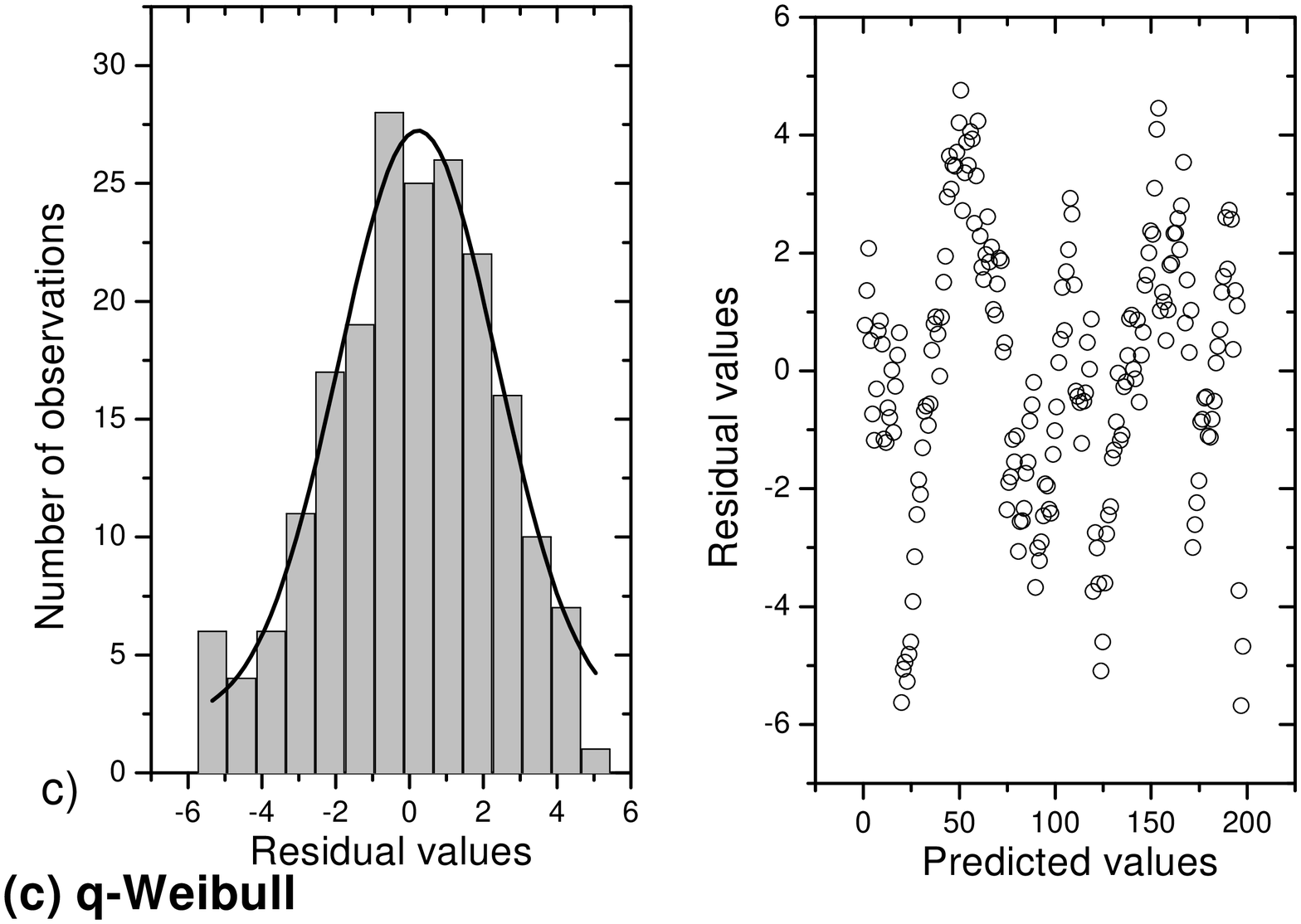}
\caption{Residual distributions to highway system when employing
{\bf a}) $q$-exponential, {\bf b}) Weibull, and {\bf c})
$q$-Weibull. The graphics on the left concern the frequency
distribution of residues and the graphics on the right refer to
predicted versus residual values. } \label{fig5}
\end{figure}
\end{widetext}
\section{Conclusion}

In several areas of nature Weibull  and Zipf-Mandelbrot
($q$-exponential with $q>1$) have an important role to describe
many  frequency distributions. For a short range, both
distributions can lead to an apparent agreement. Thus, to decide
which distribution gives a better description, it is necessary to
consider a sufficiently large range. In this work, in order to
investigate the referred controversy,  we focalized systems just
taking   a sufficiently large range of data into account.
Moreover, as to obtain a further illumination on this controversy,
we introduce a new distribution, $q$-Weibull. Such distribution
slowly interpolates the  $q$-exponential and the Weibull ones.
When either the $q$-exponential or the Weibull distribution is
able to give a good description  of a system, its respective
parameter, $r$ or $q$, is close to one. We applied $q$-Weibull
function to analyze distributions of basketball baskets, cyclone
victims, brand-name drugs by retail sales, and highway length.  We
verified that the basketball baskets distribution is well
described by a $q$-exponential with $q<1$. For cyclone victims and
brand-name drugs by retail sales, the Weibull distribution gives a
good adjustment. On the other hand, for length of highway,
$q$-Weibull distribution  gives a satisfactory adjustment. We hope
that this new distribution can be useful in other situations where
neither $q$-exponential nor Weibull ones leads to a satisfactory
result.

We thank CAPES and CNPq (Brazilian agencies)  for partial
financial support.


\end{document}